\documentclass[twocolumn,showpacs,preprintnumbers,amsmath,amssymb]{revtex4}

\usepackage{epsfig}
\usepackage{graphicx}
\usepackage{dcolumn}
\usepackage{bm}

\begin{document}

\preprint{APS/123-QED}

\title{6\,W, 1\,kHz linewidth, tunable continuous-wave near-infrared laser}

\author{Sheng-wey Chiow,$^1$ Sven Herrmann,$^1$ Holger M\"uller,$^{1,2,3}$ Steven Chu$^{1,2,3}$}

\address{$^1$ Department of Physics, Stanford University, Stanford,
California 94305-4060, U.S.A.,\\ $^2$ Department of Physics,
University of California, 366 LeConte Hall, Berkeley, California
94720-7300, U.S.A.\\ $^3$ Lawrence Berkeley National Laboratory,
One Cyclotron Road, Berkeley, California 94720, U.S.A.}

\email{swchiow@stanford.edu}


\begin{abstract} A modified Coherent 899-21 titanium sapphire laser is injection
locked to produce 6-6.5\,W of single-frequency light at 852\,nm.
After closed-loop amplitude control and frequency stabilization to
a high-finesse cavity, it delivers 4-4.5\,W with $<1\,$kHz
linewidth at the output of a single-mode fiber. The laser is
tunable from about 700-1000nm; up to 8\,W should be possible at
750-810\,nm.
\end{abstract}
\maketitle





\section{Introduction}
The improvement of tunable, high-power, low linewidth
lasers has been instrumental in advancing atomic physics.
State-of-the-art titanium:sapphire (Ti:sa) lasers span wavelengths
between 700-1000\,nm or more. Typical commercial models use
intracavity etalons for single-mode operation and achieve an
output power around 1.5-2.5\,W. Injection lock
\cite{Tanaka,Cummings,Cha} allows for dispensing with the etalons and
thus reaching a higher power of up to 5\,W so far \cite{Cha2}.
Diode lasers with tapered amplifiers, which are only available in few spectral regions, currently reach 1-2\,W (up to
2.5\,W in $\mu$s pulses \cite{Takase}) at 780\,nm and 0.5\,W near
850\,nm. Because of their lower beam quality and the need for an
optical isolator, however, they rarely deliver more than 0.5\,W
into single-mode fibers. The linewidth of both commercial single
frequency Ti:sapphire and extended-cavity diode lasers is on the
order of a few 100\,kHz. A reduction to the kHz and even Hz level
can be achieved by stabilization to high-finesse optical cavities.
Here, we report an injection-locked Ti:sapphire laser system that
achieves up to 6.5\,W at 850\,nm, up to 4.5\,W after a
single-mode, polarization maintaining (SM-PM) fiber, and $<1\,$kHz
linewidth by stabilization to a high-finesse reference cavity.

\section{Description}
Figure 1 shows the setup. It can be roughly divided into the
injection lock system shown on the left, the frequency
stabilization in the upper right, and the amplitude control shown
in the lower right part of the figure.

\begin{figure}
\centering\epsfig{file=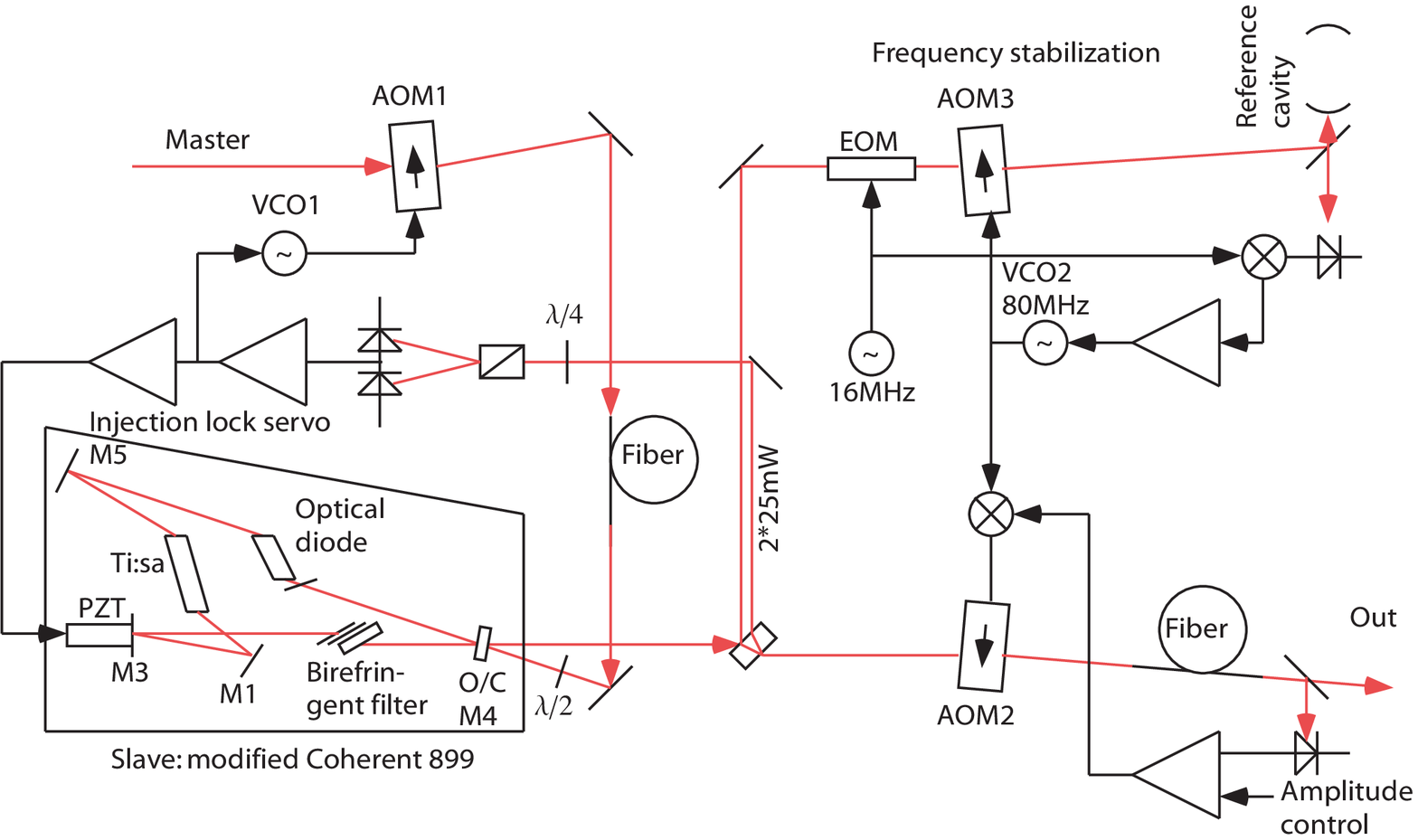, width=0.5\textwidth}
\caption{Setup.}
\end{figure}

The master laser (not shown) is a Coherent 899 with 1.2-1.3\,W
output power when pumped with 10\,W from a Coherent Verdi V-10. It
is frequency stabilized using the internal piezoelectric
transducer (PZT) and Brewster plate as actuators. For this
purpose, we modified the control box as described in Ref.
\cite{Haubrich}. As a frequency reference, we use an
extended-cavity diode laser (ECDL) \cite{tracked}, which is in
turn stabilized to a $D_2$ transition in a cesium vapor cell.
A detuning of 0 to $\pm20$\,GHz can be set by a microwave
synthesizer.

The master passes an acousto-optic modulator, AOM 1 (Fig. 1),
driven by voltage controlled oscillator VCO 1 at about 120\,MHz.
The light is then coupled to an SM-PM fiber, in order to decouple
the alignments of the master and slave. About 0.8\,W are typical
after the fiber.

The slave laser is another Coherent 899. To increase the output
power, we removed the thick and the thin etalon as well as the
Brewster plate, as they are not required for injection locked
operation \cite{birfilter} and replaced the output coupler by one
having 10\% transmission \cite{coupler}. Pumped with 19\,W from a
Coherent Innova 400 argon-ion laser (not shown), we obtain 5.5\,W
output from the free-running slave.

For injection lock, the passive resonance frequency of the slave
laser's cavity must be stabilized to the master. For generating an
error signal without adding sidebands to the radiation, we use the
H\"ansch-Couillaud method \cite{Couillaud}. The master laser's
polarization has a small angle relative to the slave, set by a
half-wave retardation plate. The small angle means that most of
the electric field of the master can add to the slave's field (at
lower master power, an angle of 45$^\circ$ would be used to give a
larger error signal). To generate the error signal, a small
fraction (25\,mW) of the output beam is picked up and its
polarization analyzed.

The error signal is fed back to the piezoelectric transducer (PZT)
that translates the lower fold mirror M3 \cite{remark}. We found
the $\sim 5$\,kHz bandwidth of this path to be sufficient for
locking, but only just. Therefore, we add a fast (about 100\,kHz
bandwidth) feedback channel to VCO 1 (mini-circuits POS-150), in
analogy to \cite{Cummings,Cha}. This makes the lock reliable, at
the price of some broadening of the master laser's linewidth. With
up to 0.8\,W from the master laser, there was no need to optimize
for low master laser powers. We confirmed, however, that the slave
can still be locked with about 200\,mW from the master. Previous
work used master laser powers of 1\% \cite{Tanaka} to 1.5\%
\cite{Cummings,Cha} of the slave, which would be in the range of
50-100\,mW for our power level.

The combined power of master and slave (6\,W typical with a slope efficiency of $\sim 40$\%; 6.75\,W have
been achieved \cite{pumppower}) is deflected by AOM 2 (Crystal
Technology 3080-122) and coupled into a short \cite{SBS} SM-PM
fiber (OFR). In our application, the AOM is
usually driven by 100-$\mu$s pulses; nevertheless, we confirmed
that cw operation for at least a few h is feasible without
destroying the fiber. An adjustable OFR fiber port is used for the spatial mode-matching of the injection and the slave laser cavity: The mode matching is optimized so that the highest coupling efficiency of the combined light into the short fiber is obtained.

Since the injection lock leads to a fixed phase relationship
between the master and the slave, given suitable alignment, the
outputs of both lasers can be added coherently: we obtain 4\,W
typical (4.5\,W best) after the fiber; this means that one of the
AOM or the fiber is at least 81\% efficient. This efficiency is
similar to what we obtain with either the master or the slave
alone, confirming that the outputs of the two lasers do indeed add
coherently, so that the power of the combined lasers into the mode
of the fiber is close to the sum of the individual ones.

The acoustical delay of the AOM is reduced to about $0.5\,\mu$s by
moving the transducer close to the beam with a translation stage;
this adjustment must be made with care, as the very strong
Ti:sapphire beam instantly destroys the AOM if it ever hits the
transducer.

The linewidth of the locked slave is about 500\,kHz full width at
half maximum. To reduce it, we split off a small sample, shift it
by AOM 3 and stabilize it to a Fabry-Perot cavity with a linewidth
of about 25\,kHz (finesse 75,000). The error signal is generated
by the Pound-Drever-Hall method. Phase modulation at 16\,MHz is
applied by an electro-optical modulator (EOM). The light reflected
by the cavity is detected and down-converted to obtain the error
signal. We feed back to AOM3's frequency via VCO 2, with a loop
bandwidth of about 300\,kHz.  VCO 2 also controls the main beam
frequency by driving AOM 2. In effect, the main beam is stabilized
to the cavity independent of the intensity setting of AOM 2. AOM 3 is positioned with a translation stage to match the delays of AOM 2 and AOM 3 so that the noise in the beat note measurement (will be described in the following paragraphs) is minimized. A
slow (several 100\,Hz bandwidth) servo (not shown) zeros the dc
signal to the VCO by feeding back to the cavity via a PZT. This is
used to take out the cavity drift and means that on long time
scales, the slave is locked to the cesium spectroscopy.

To verify the linewidth of the system, a beat note measurement
with an independent low-linewidth reference laser would be ideal.
Such a laser was unfortunately not available to us. As the second
best option, we use the {\em transmitted} light of the cavity as a
reference. This method is as good as an independent laser for
noise frequencies much higher than the cavity linewidth: Such
noise components are filtered out by the cavity and are thus
suppressed in its transmission. Regardless of frequency, the
method measures the laser frequency relative to the cavity
resonance. To minimize fluctuations of the cavity resonance
itself, the cavity is held by viton rings inside a hermetically
sealed (but not evacuated) stainless steel chamber to provide
acoustic and thermal insulation.

For the beat note measurement, we overlap a sample of the power
after the fiber with the transmitted light of the cavity on a
photodetector. The measured spectrum (Fig. 2) reveals a linewidth
of 1.3\,kHz (full width at half maximum), limited by the 1\,kHz
resolution of our spectrum analyzer. Assuming that the linewidths
add geometrically, we conclude that the laser linewidth is
800\,Hz. Frequency modulation sidebands at 24\,kHz are an effect
of mechanical vibrations of the cavity. The wideband spectrum
(Fig. 2, right) shows an about 800\,kHz wide noise pedestal on the
order of -80\,dBc/Hz, the residual of the free-running laser
noise. At frequencies more than 400\,kHz from the carrier, the
noise drops rapidly.

\begin{figure}
\centering\includegraphics[width=7cm]{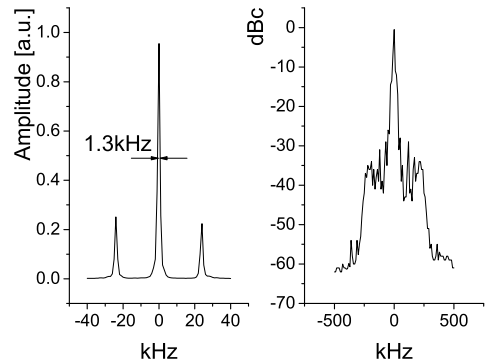} \caption{Beat
measurement between the fiber output and the transmission of the
cavity. Left: Linear scale of the innermost 80\,kHz to show
linewidth. 1\,kHz resolution. Right: 1\,MHz scan shown with
logarithmic scale in dB referred to carrier. 10\,kHz resolution
bandwidth.}
\end{figure}

The laser system as installed is tunable within 780-925\,nm, and
more than 700-1020nm with the short or long wavelength optics
sets. Between 750-810\,nm, the output power of Ti:sapphire lasers
such as the Coherent 899 is typically about 30\% higher than at
850\,nm, so about 8\,W are possible. We have, however, not
ascertained that fiber coupling of this is feasible. If wide
tunability is not required, a considerable cost reduction can be
achieved by using a diode laser (possibly with a tapered
amplifier) as the master laser.

\section{Conclusion}
In summary, we have presented an injection locked laser system.
The master and slave laser's power add coherently to up to
6.75\,W. The system produces power- (up to 4.5\,W) and frequency
stabilized (linewidth below 1\,kHz) and mode-filtered radiation at
the output of a single-mode, polarization maintaining fiber. The
optical frequency is locked to a Cs hyperfine transition with an
offset between 0 and $\pm 20$\,GHz; a high-finesse cavity is used
for linewidth reduction. The amplitude stabilization with its
250\,kHz loop bandwidth allows us to precisely shape light pulses
of arbitrary form. This system is ideally suitable for demanding
applications such as high-order Bragg diffraction \cite{BraggPRL}
in atom interferometry, or as a basis for a powerful, tunable uv
source by frequency multiplication \cite{Mes}.

\subsection{Acknowledgments} H.M. and S.H. thank the Alexander von Humboldt Foundation. This
material is based upon work supported by the National Science
Foundation under Grant No. 0400866 and by the Air Force Office of
Scientific Research under Award Number FA9550-04-1-0040.


\begin{thebibliography}{99}
\bibitem{birfilter} We also tested
removing the optical diode and the birefringent filter. We found,
however, that this makes the laser exceedingly difficult to align
and thus we achieve higher power with these elements in place.
\bibitem{remark} It is important to limit the PZT volatge to below the maximum of 500\,V
(using Z-diodes, for example).
\bibitem{pumppower} Increased pump power can
increase the total power to a maximum of 7\,W, but also results in
instable operation. Eventually, at more than 22\,W pump, the
output power drops. We attribute this behavior to thermal lensing.
\end{thebibliography}
\end{document}